\documentclass[sigconf]{acmart}

\AtBeginDocument{%
  \providecommand\BibTeX{{%
    \normalfont B\kern-0.5em{\scshape i\kern-0.25em b}\kern-0.8em\TeX}}}

\setcopyright{acmcopyright}

\copyrightyear{2020}
\acmYear{2020}
\acmDOI{10.1145/3383313.3412255}

\acmConference[RecSys '20]{Fourteenth ACM Conference on Recommender Systems}{September 22--26, 2020}{Virtual Event, Brazil}
\acmBooktitle{Fourteenth ACM Conference on Recommender Systems (RecSys '20), September 22--26, 2020, Virtual Event, Brazil}
\acmPrice{15.00}
\acmISBN{978-1-4503-7583-2/20/09}

\usepackage{amsmath}
\usepackage{subcaption}
\usepackage{graphicx}
\usepackage{xcolor,colortbl}
\definecolor{mygray}{gray}{0.85}
\usepackage{multirow}

\usepackage{amsmath}
\DeclareMathOperator*{\argmax}{arg\,max}

\begin{document}

\title{Goal-driven Command Recommendations for Analysts}

\author{Samarth Aggarwal}
\authornote{Both authors contributed equally to this research. The work was done when all the authors were affiliated with Adobe Research.}
\affiliation{%
  \institution{Indian Institute of Technology Delhi}
  \city{Delhi}
  \country{India}}
\email{cs1160395@cse.iitd.ac.in}
\author{Rohin Garg}
\authornotemark[1]
\affiliation{%
 \institution{Indian Institute of Technology Kanpur}
 \city{Kanpur}
 \country{India}}
\email{sronin@iitk.ac.in}

\author{Abhilasha Sancheti}
\affiliation{%
  \institution{Adobe Research}
  \city{Bangalore}
  \country{India}}
\affiliation{%
  \institution{University of Maryland}
  \city{College Park}
  \country{USA}}
\email{sancheti@adobe.com, sancheti@cs.umd.edu}

\author{Bhanu Prakash Reddy Guda}
\affiliation{%
  \institution{Adobe Research}
  \city{Bangalore}
  \country{India}}
\email{guda@adobe.com}

\author{Iftikhar Ahamath Burhanuddin}
\affiliation{%
 \institution{Adobe Research}
  \city{Bangalore}
  \country{India}}
\email{burhanud@adobe.com}

\renewcommand{\shortauthors}{Aggarwal, Garg, Sancheti, Guda and Burhanuddin}

\begin{abstract}
  Recent times have seen data analytics software applications become an integral part of the decision-making process of analysts. The users of these software applications generate a vast amount of unstructured log data. These logs contain clues to the user's goals, which traditional recommender systems may find difficult to model implicitly from the log data. With this assumption, we would like to assist the analytics process of a user through command recommendations. We categorize the commands into software and data categories based on their purpose to fulfill the task at hand. On the premise that the sequence of commands leading up to a data command is a good predictor of the latter, we design, develop, and validate various sequence modeling techniques. 
  In this paper, we propose a framework to provide goal-driven {\emph{data command}} recommendations to the user by leveraging unstructured logs. 
  We use the log data of a web-based analytics software to train our neural network models and quantify their performance, in comparison to relevant and competitive baselines. We propose a custom loss function to tailor the recommended data commands according to the goal information provided exogenously. 
  We also propose an evaluation metric that captures the degree of goal orientation of the recommendations. 
  We demonstrate the promise of our approach by evaluating the models with the proposed metric and showcasing the robustness of our models in the case of adversarial examples, where the user activity is misaligned with selected goal, through offline evaluation. 
\end{abstract}




\maketitle


\section{Introduction}
There has been tremendous growth in the domain of data analysis as the volume of data has increased, and as the capabilities to support processing of this data have advanced \cite{mckinsey}. Analysts, and more generally, users of data-centric software applications, need to make several selections within the software application to achieve certain objectives, gather insights from the data and make decisions \cite{selene2014review}. Considering the sheer volume of data to be analysed, there is now a demand on systems to query, analyze and draw inferences with a low latency. This demand also carries over to users assigned with the task of analysing the data. Recommender systems have been used time and again in a variety of different applications to guide the users. They solve two major concerns:
\begin{enumerate}
    \item When a user is faced with a raft of different options to choose from, recommender systems act as a primary filter for options that are completely irrelevant. This leaves the user with a choice among a relatively smaller number of options \cite{resnick1997recommender}.
    \item When a novice user lacks the skills and knowledge to choose from the different options provided, recommender systems act as a guide for the user in making a selection \cite{blandford}. We refer to this click as a {\emph{command}} or {\emph{action}} that gets registered in the log data when a user interacts with the interface of the system.
\end{enumerate}

The domain of analytics offers both these problems simultaneously, and our solution approach proposes to build a system that caters to both of them. Introducing recommendations to help the analysts decide what activity to perform and where to look in the data to discover insights can boost productivity. However, recommendations can be distracting if they are irrelevant to the analyst's goal and do not provide guidance towards the insights that the analyst needs. 

\begin{figure*}[ht]
    \centering
    \includegraphics[width=1.0\linewidth]{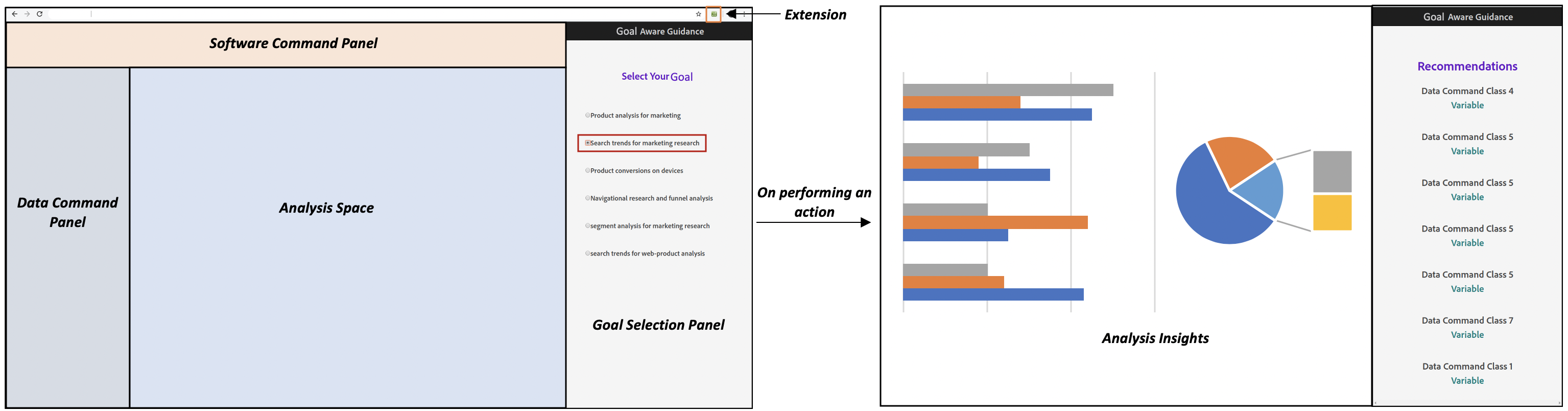}
    \caption{\small{Left: The landing page of a user on logging into the analytics product. The \emph{goal selection panel} contains the descriptions of the goals from which the user selects. Right: The \emph{data command} recommendations provided by our model. Transition: On clicking a \emph{software} or \emph{data command}.}}
    \label{fig:final_ui}
    \vspace{-4mm}
\end{figure*}

The interaction mechanism illustrated in this paper borrows the notion of goal and applies it to software assistance that is provided by a recommender system (Fig. \ref{fig:final_ui}). Specifically, the goal information provided as input by the user during a session of interaction with a software application guides the recommendations the user receives. Our study is under the purview of an interactive data analysis and visualization software for web analytics. 
We believe that claims and results presented here generalize to assistance in education and other data-centric domains where the notion of goals can be formalized. Our contributions in this work are summarized below. 
\begin{enumerate}
\item  We conceptualize the notion of a goal in data analytics software applications and propose models where goal information is provided exogenously. We design, develop and evaluate sequence-based goal-driven models, which are better predictors of \textit{data commands} compared to traditional recommender systems that were unaware of the user's goal. We demonstrate the performance of our approach through offline evaluations.
\item  We design a custom loss function for models which steer the recommendations towards the user's goal. The loss function involves a probability distribution component, which is different for each of the identified goals, thereby generating an ensemble of fine-tuned models. 
\item We also introduce a measure to evaluate the degree of goal orientation in recommendations provided by the proposed models.
\end{enumerate}

To demonstrate the robustness of our model, we report the performance of our models on adversarial examples. The results showcase how our models would operate in cases of misspecified information where user's activity is misaligned with provided goal information.

The paper is organized as follows. We present related work in Section \ref{relatedwork}. In Section \ref{dataset}, we describe the dataset used in our experiments. Sections \ref{models} and \ref{experiments} describe our goal-driven models and baselines, and experimental results respectively. Next in Section \ref{evaluation}, evaluation metrics and results are showcased.

\section{Related Work}
\label{relatedwork}



\vspace{2mm}
\noindent \textbf{Goal-driven recommendations:}

There is a vast amount of literature related to the concept of goals in process mining, web mining, education \cite{zhou:task-goal, jiang:goal1, jiang:goal2} and HCI \cite{card1983psychology}. In the paper \cite{Lumiere}, a {\emph{goal}} is defined as {\emph{a set of target tasks or subtasks at the focus of a user's attention}}. We have applied this definition to our setting, where the user selects the target task at the start of the analysis session, tasks in turn are modeled as topics, and recommendations are guided by the task selections.

While a significant amount of research has been carried out to understand and analyze patterns in user behavior from application log data \cite{FrequentTasks, Workflows, LogAnalysis, davison1998predicting, matejka2009communitycommands}, limited research has been published to infer goals from usage logs and explicitly incorporate goal information to recommend future commands \cite{React, RecommendWorkflows, CAPA, nambhi2019stuck}. 
The paper \cite{Eventlogs} models the workflows in event log data based on probabilistic suffix trees (PST) and \cite{Workflows, RecommendWorkflows} deploy the techniques of topic modeling to determine the workflows or tasks. In \cite{RecommendWorkflows}, a system to recommend video tutorials based on the command logs from the user activity is discussed. Here a hierarchical approach that operates both at the task and command level is proposed. The authors consider topics determined through a topic model as tasks and at the command level they utilize frequent pattern mining and itemset mining techniques to extract frequent command patterns for videos. The paper \cite{RecommendWorkflows} builds its approach on top of a study of frequent user tasks from product log data \cite{FrequentTasks}. Additionally, they leverage topic modeling as an antecedent layer to capture diverse behavior patterns.  This modification has allowed them to represent command patterns from less frequent tasks more faithfully. 
Elsewhere  the concept of context is captured by representing the sessions through trees \cite{React}. 

The paper \cite{Workflows} explored the problem in the same direction as \cite{RecommendWorkflows}. Unlike the latter approach, they intend to model the workflows using probabilistic suffix trees. In a similar vein, \cite{nambhi2019stuck} have proposed neural network architectures \emph{TaskRNN} and \emph{JTC-RNN} to predict the next command in the sequence. Their architecture is similar to that of \emph{TopicRNN} \cite{TopicRNN} from the domain of text analysis. 
In line with prior work \cite{RecommendWorkflows, nambhi2019stuck, Workflows}, we use topic modeling techniques to model goals that users carry out by executing a sequence of commands. As topic modeling techniques such as Latent Dirichlet Allocation (LDA) \cite{LDA} suffer from data sparsity issues, we have used Biterm Topic Modeling (BTM) \cite{BTM} to identify goals from the log data as typical user sessions are small in size. We define the goals mathematically based on the outputs of the BTM model. Researchers have proposed architectures on incorporating the document context in the form of topic-model like architecture, thereby granting a broader document context outside of the current sentence \cite{TDLM}. Inspired by this language modeling architecture, we propose methods to incorporate current goal information for predicting the next \emph{data command} in the sequence. 
One of the major drawback of these approaches is that they utilize standard cross-entropy loss function which does not consider any goal information while penalizing their models. To the best of our knowledge, there has also been no work which introduces a loss function to steer the recommendations based on the goal under consideration.

\vspace{2mm}
\noindent \textbf{The Fine-tuning paradigm:}
Our proposed models employ the technique of fine-tuning inspired by the Generative Pre-Training (GPT) \cite{radford2018improving}, Generative Pre-Training 2 (GPT-2) \cite{radford2019language} modes of training. 
Both GPT and GTP-2 models demonstrated large gains on natural language understanding tasks by pre-training of language models on a diverse corpus of unlabeled text, followed by discriminative fine-tuning on each specific task. The GPT and GPT-2 models utilize task-aware inputs during fine-tuning to achieve effective transfer while requiring minimal changes to the model architecture. In this paper we apply a similar framework for the first time to provide goal-driven recommendations by initializing parameters from a trained model for generalized recommendations and fine-tuning with custom loss function to achieve goal specific models. This is very effective in providing recommendations where the user activity is misaligned with the goal given as input.

\section{Dataset}
\label{dataset}
The application under consideration is a web-analytics system used to track, report, analyze, and visualize web traffic. Traces of user activity within this software is available as log data. This log data contains the user interface clicks that are captured while the user interacts with the system. We term these clicks as commands, and they fall into the following two categories (1) Software Commands; and (2) Data Commands. 
The first category of commands is related to analyzing data by the user through mental effort via tables and visualizations, or by applying software tools with underlying statistical or machine learning models. This set of commands is denoted by SC for {\emph{Software Commands}}. The commands which specify the data to be analyzed fall into the second category. This second set of commands is denoted by DC for {\emph{{Data Commands}}}. 
Each of the data commands can be split into a class and corresponding variable. We observe that selection of these categories of commands is iterative in nature. We would like to provide examples from a familiar framework --- spreadsheet applications. 
In spreadsheets, the software commands refer to opening, loading or saving a spreadsheet, changing the color of a cell, whereas data commands include making column selections, sorting a data column in an increasing or  decreasing order, computing aggregation functions such as sum or maximum or minimum of a data column. 
In this example the class of data commands corresponds to sorting and the variable is the value on which the data is sorted. We consider the data command class along with the variable, to be a unique data command for the rest of the paper. 

Often, an analyst begins analysis with the goal to provide explanations for certain patterns which are captured in the data.
The software commands merely facilitate data analysis. On the other hand, the answers sought by the analyst are closely related to the data that is being analyzed. Our objective is to build a system to assist the user by providing guidance on where to look at in the data. Therefore, out of these two sets of commands, we would like to predict only the data commands. 
After pre-processing three months of usage logs from April 2018 to June 2018, we have extracted data for a few hundred users in the specified duration. 
The logs with sparse user activity were dropped. The extracted logs were then split into sessions\footnote{A session is a sequence of commands executed by a single user from entering the application to exiting the application.} considering the \emph{sign-out} software command as the delimiter.
To handle the cases where the user does not end the session explicitly by issuing the \emph{sign-out} software command, we have considered six hours of inactivity as a signal for the end of the session. The duration of inactivity we chose is well beyond the time after which the application terminates the session automatically, if no user response is recorded. The session thus captured were $55K$ in number. 
There were around $838$ unique commands, which were obtained after dropping the commands that were logged to indicate user interface events and not explicitly executed by the users. Out of these $838$ commands, $766$ commands were the data commands (DC), and the rest of them were the software commands (SC). There is a limited amount of our subjective judgement to arrive at this categorization. 
\begin{figure}[h!]
  \[
   \boxed{SC, DC, DC, SC, ..., SC,}\ SC, SC, \textbf{DC}, DC, SC, ..
  \]
  \caption{\small{In this example session of user activity, sequence data is generated for model training by placing a window at each position in the session. The next command that needs to be predicted for the sequence in the window is the immediate data command after the window ends (command in bold).}}\vspace{-2mm}
  \label{fig:sequence_diag}
\end{figure}

After the pre-processing phase, the average number of commands per session was found to be $30$. For uniformity, sessions with length less than $30$ were dropped, 
and sessions with length greater than $30$ were handled based on sliding window approach which is explained in Figure \ref{fig:sequence_diag}. 
A total of $2.7M$ sequences, thus generated, were split into training, validation and test sets on the basis of sessions, with $75:12.5:12.5$ split for the three sets respectively. 
\section{Goal-Driven Models}
\label{models}

The interaction in our proposed system has the user selecting goals represented by phrases at the start of a session in line with their intended task. We reiterate that the validation of this proposal is through offline evaluation as described in Section \ref{evaluation}. Example phrases for goals for the dataset under consideration are also provided in that section as well as in Figure \ref{fig:final_ui}.

To recommend data commands based on the goal information provided by the user, we propose variants of sequence to sequence (seq2seq) models which are compared against more traditional approaches such as popularity based models, bag-of-commands models, and Markov models. The goal information is delivered to the seq2seq models using two approaches by (1) providing training data that corresponds to each specific goal, and (2) by building goal informed models. However, before recommending the data commands, the first sub-problem is to identify the possible goals from the user log data. 

 
\subsection{Goal Identification}
\label{btm}
The problem of making recommendations goal-aware requires solving two problems, goal identification and adding goal information to the recommender system. The former is a pre-requisite for the latter and was done from user log data, though does not affect the latter.

\begin{figure*}[h!]
    \centering
    \scalebox{0.7}{
    \includegraphics[width=1.0\textwidth]{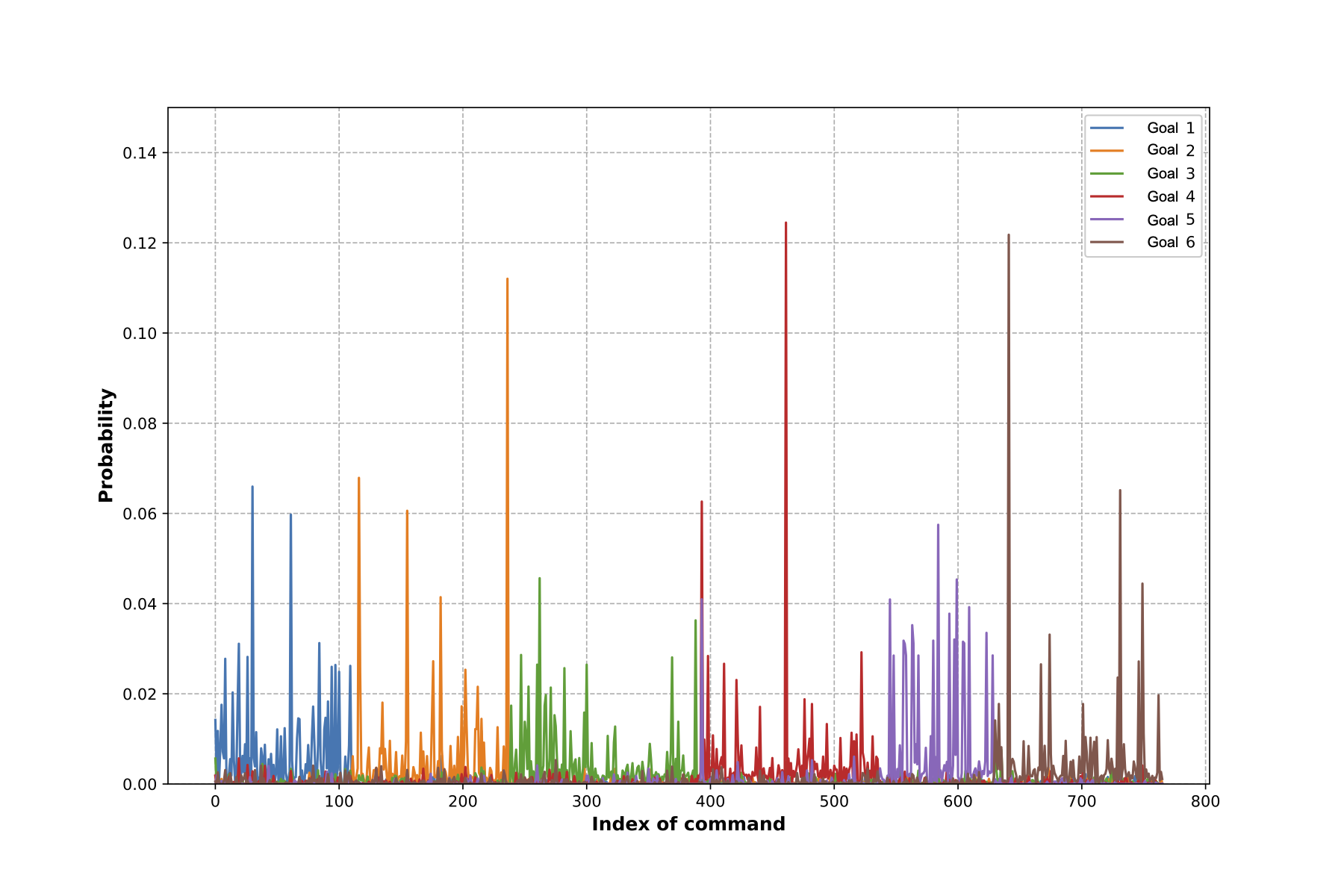}}
    \caption{The probability distribution of commands for each of the identified $K=6$ goals: $P(dc\vert goal)$. The commands are grouped together goal-wise. This graph illustrates the distinguishable definitions of goals through their probability distributions.}
    \label{fig:prob_dist}
\end{figure*}

In this section, we draw an analogy between a collection of natural language documents and a collection of command sequences. A word, which is a fundamental unit of a document corresponds to a command in a sequence. Unlike words, in our setting there are two categories of commands, data commands ($DC$) and software commands ($SC$) as described in Section \ref{dataset}. 
The words constitute the documents and similarly the commands ($DC \cup SC $) form the sequences.
The command distribution $\phi_G$, obtained from the BTM \cite{BTM} model, contains the probability values for all the commands, software and data commands. Each topic output of the BTM model corresponds to a goal $G$. We use BTM, instead of using more popular approaches like LDA, to alleviate the data sparsity problem. This problem arises due to co-occurrence matrix for each and every pair of commands being very sparse. As we aim to predict only the data commands, those probabilities that pertain to the data commands are considered and normalized to generate a distribution for the goal $G$. 
\begin{equation}
    \label{equ:intent_def}
    P(dc = dc_{i} \vert goal = G) = \frac{\phi_{G}[dc_{i}]}{\sum_{c\in DC}\phi_G[c]}
\end{equation}
where ${dc}_i\in \textit{DC}$ is the $i^{\textrm{th}}$ data command, $\phi_G[c]$ denotes the probability of the command $c$ in the command distribution for the goal $G$. The computed distribution, $P(dc \vert goal = G)$, is considered as the definition for the goal $G$ (Fig. \ref{fig:prob_dist}). 

\subsection{Goal Coherence}
\label{coherence}
To decide upon the correct number of goals, we utilize standard coherence evaluation metrics, explained in this section. 
As an early evaluation method of topic models, \cite{alsumait2009topic} defines an unsupervised goal ranking measure based on three prototypes of irrelevant and insignificant topics. 
Then, a topic significance score is computed by applying various similarity measures such as cosine, correlation and dissimilarity measures such as KL divergence \cite{kullback1997information} to these three prototypes. It is unclear to what extent their unsupervised approach and objective function agrees with human judgements, however, as they present no user evaluations \cite{newman2010evaluating}.
Consequently, more and more works have been done to correlate the evaluation measure with human judgement. 
Two such state-of-the-art goal coherence (topic coherence) evaluation measures are the UCI and UMass. Both the measures compute scores for all pairs of data commands in a goal cluster. They differ in defining the score for a pair of data commands. 

The UCI measure \cite{newman2010evaluating} uses pairwise score function, Pointwise Mutual Information (PMI) to compute the score between two data commands. The UMass measure \cite{mimno2011optimizing} uses a pairwise score function similar to UCI.
\begin{align}
    score_{\text{\scriptsize{\emph{UCI}}}}(dc_i, dc_j) = \log \frac{p(dc_i, dc_j)}{p(dc_i)p(dc_j)} \\
    score_\text{\scriptsize{\emph{UMass}}}(dc_i, dc_j) = \log \frac{M(dc_i, dc_j) + 1}{M(dc_i)}
\end{align}
where $p(dc_i)$ represents the probability of seeing data command $dc_i$ in a session, and $p(dc_i,dc_j)$ is the probability of observing both $dc_i$ and $dc_j$ co-occurring in a session computed as follows
\begin{equation}
    \label{uciprobs}
    p(dc_i)=\frac{M(dc_i)}{M}\ \ \textrm{and}\ \ p(dc_i,dc_j)=\frac{M(dc_i, dc_j)}{M}
\end{equation}
where $M(dc_i)$ is the count of sessions containing the command $dc_i$, $M(dc_i,dc_j)$ is the count of sessions containing both commands $dc_i$ and $dc_j$, and $M$ is the total number or sessions. The UCI score for a goal, $G$, is computed as $\textrm{mean}\{score_\text{\scriptsize{\emph{UCI}}}
(dc_i, dc_j), dc_i, dc_j\in G, i\ne j\}$. 
The overall UCI score equals the mean UCI score of all the goals. The overall UCI score for a model with $t$ goals, $CS_\text{\scriptsize{\emph{UCI}}}(t)$ equals the mean UCI score across those $t$ goals.


The overall UMass score, $CS_\text{\scriptsize{\emph{UMass}}}$, is computed similar to the UCI measure. 
Higher UCI and UMass scores indicate better groupings, since if two data commands in a goal really belong together we would expect them to show up together very frequently. The process of choosing optimal number of goals for our data by considering both these measures is discussed in Section \ref{experiments}.
\subsection{The Ensemble Approach}
\label{ismodels}
The information of goal can be implicitly incorporated into the models through the data that is provided as input to them.
Several applications have used this approach earlier to generate models specific to a data distribution observed in a particular class or task in general. Multi-class classification \cite{Mayoraz1999SupportVM, Rifkin2004InDO, liu2005one, polat2009novel}, is a classic example of this approach where the data distribution of a class is assimilated through an ensemble of models. Similar to this approach, we propose one model per goal, which is explicitly trained to model the distribution of the data of that particular goal. The technicalities of the model are as follows. 
Our proposed models use multi-layered Long Short-Term Memory (LSTM) \cite{LSTM} to encode the input sequences of commands into vectors of fixed dimensionality. Given a sequence $S$ with $c_i\in \text{DC} \cup \text{SC}, i\ \text{in}\ [0,L]$ commands, we first embed the commands through an embedding matrix. This sequence of embedded commands is then provided as an input to an LSTM encoder which computes representations of the commands by summarizing the information. The representation from the last time step of the LSTM hidden unit, denotes the semantic representation of the command sequence in the latent space. It is then fed as input to a fully connected layer, followed by a softmax layer \cite{bishop2006pattern} for predicting the succeeding data command in the given sequence.
Mathematically, at each step of data command prediction, the following probabilities are computed to generate next data command in the sequence $S$:
\begin{equation}
\label{equ:prob}
    Pr(\hat{dc} = {dc}_i \vert {c}_0, {c}_1, \dots, {c}_L)
\end{equation}

$K$ such models are trained, one for each of the $K$ goals identified through the BTM model. 
The commands which are model predictions are the ones which have the top probabilities.
\begin{equation}
\label{equ:prob-top-n}
    \argmax_{dc_{i} \in \textrm{DC}} Pr(\hat{dc} = {dc}_i \vert {c}_0, {c}_1, \dots, {c}_L)
\end{equation}

\cite{Gehring2017ConvolutionalST} have shown the tantamount performance of convolutional neural networks (CNN) compared to recurrent neural networks (RNN) in sequence modeling. 
Compared to recurrent models, computations over all elements can be fully parallelized during training facilitating us to better exploit the GPU hardware and optimization. This is due to two reasons (1) the fixed number of non-linearities in the model architecture and (2) the independence between the number of non-linearities and the length of input command sequence. Consequently, we have also explored the space of CNN's in modeling the command sequences.

We utilize the convolution operation with max-over-time pooling operation layer \cite{kim2014convolutional, Collobert2011NaturalLP}, to generate embedding representations for the command sequences. The motivation behind this operation is to capture the most significant feature---one with the highest value---for each feature map. The max-over-time pooling operation also handles the problem of having variable length command sequences. The features obtained from the convolutional filters, are concatenated to produce a single representation of the command sequence. This representation is further processed through dense and softmax layers for predicting the next data command in the given sequence. The mathematical representation of this model remains the same as shown in formula \ref{equ:prob}. 

\subsection{Goal Informed Models}
\label{iimodels}
In this section, we propose models that are provided with goal information explicitly. While ensemble of classifiers method is a competitive baseline for multi-class classification task, this approach of providing the goal information implicitly fails in general while predicting data commands for sequences which do not align with the goal. Consequently, we impose additional constraint on the seq2seq model to capture the relation between goal and the command sequence by providing the goal information explicitly \cite{TDLM, nambhi2019stuck}. This approach has the advantage of training the model on relatively larger set of data thereby resulting in better accuracy \cite{banko2001scaling}. We experiment with three ways of incorporating the goal information into the models each of which perform in par with the other two models.   

The first two sets of experiments use the one-hot representation to provide goal information to the models. The reasons why we chose one-hot representation for the goals are two-fold. (1) It is desired that the goals should have distinguishable representations, and one-hot representation distinguishes each goal in the goal vocabulary from every other goal. 
(2) The goals should not be prioritized in terms of representations when a global model is trained on data points from all the goals. 
Our first proposed model, Goal Concatenated Representation (\emph{GCoRe}), concatenates the one-hot representation ($1_G$) of the goal to the effective representation of the input command sequence, obtained after the LSTM encoder layer. The second variant, Goal Concatenated Commands (\emph{GComm}), provides the goal information before the LSTM layer. It is expected from this model to incorporate the goal information while representing the input sequences and thereby provide better results. The one-hot representation of the goal is concatenated to each of the embeddings of the commands in the sequence and is fed as input to the LSTM encoder. 


\begin{figure*}[h!]
    \centering
    \includegraphics[width=1.0\linewidth]{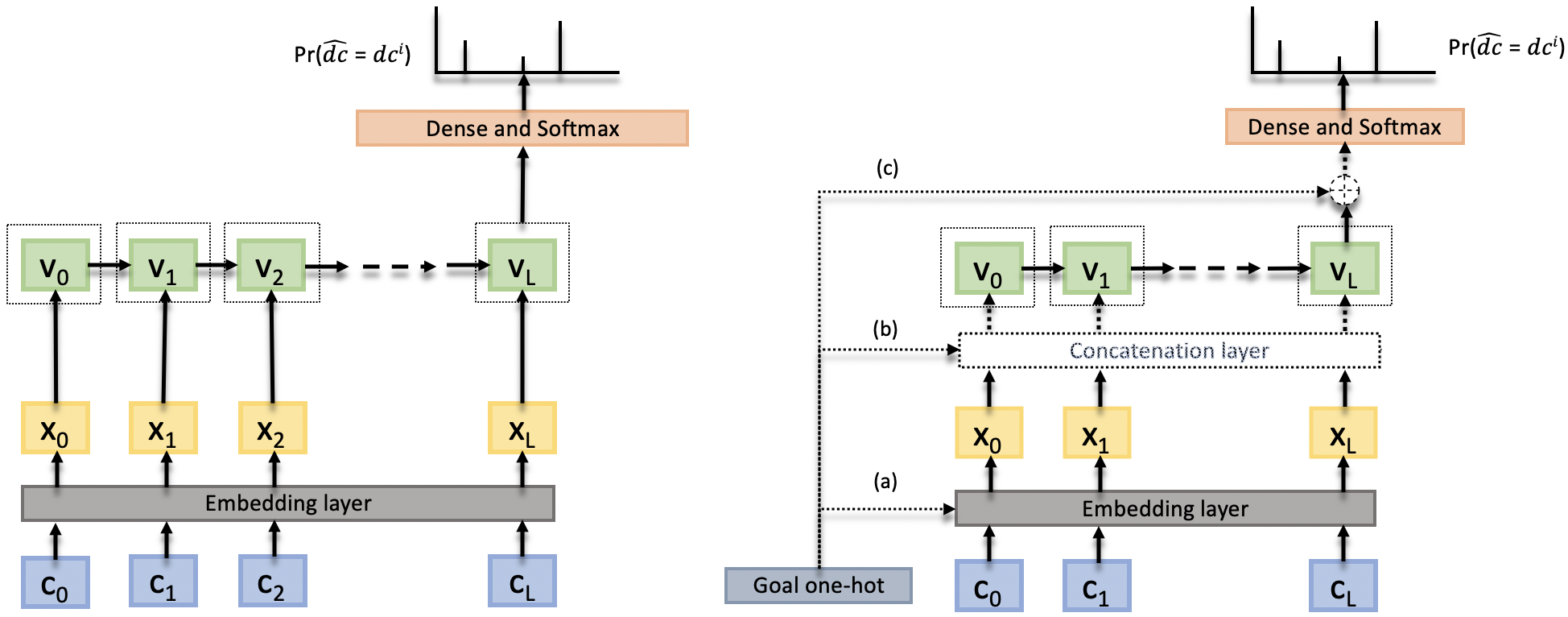}
    \caption{Left: The architecture diagram for the vanilla LSTM model. Right: The architecture diagram for goal informed (a) \emph{GAIn} (b) \emph{GComm} and (c) \emph{GCoRe} models.}
    \label{fig:incore_incomm_archi}
\end{figure*}

The third proposed model, Goal Appended Inputs (\emph{GAIn}), assumes the goals and the commands to be in the same latent representation space \cite{djuric2015hierarchical}. The goal information is provided as input at the first time step of the LSTM unit similar to the proposed architecture in \cite{vinyals2015show}. The goal and command embeddings are trained together in the same $k$-dimensional space to generate cluster representative embeddings for the goals. 
The embedding matrix is modified to facilitate the embeddings for the goals. The goal is prepended to command sequence and is fed as input to the model. The modified input sequence to the LSTM will be $(G, c_0, c_1, \dots, c_L)$. 
The mathematical representation of the above proposed models is given by:
\begin{equation}
\label{equ:prob_2}
    Pr(\hat{dc} = {dc}_i \vert {c}_0, {c}_1, \dots, {c}_L, G)
\end{equation}
Here, $G$ is the goal assigned to the command sequence $(c_0,$ $c_1,$ $\dots,$ $c_L)$ by BTM. The modifications can be applied to the CNN models as well, with the rest of the computations remaining intact. 
\subsection{Loss Function}
So far the various proposed models are variations of next data command prediction models based on seq2seq modeling. 
The standard cross-entropy loss function $\mathcal{L}_{CE}(\theta)$ \cite{janocha2017loss}, when applied to these models, makes sure that the recommended data commands are actually aligned with the input sequence. 
However, a severe limitation to this standard loss function in our setting is that this loss function does not consider the goal orientation while penalizing the models. 
These models do not have access to crucial information provided by the BTM, which is the definition of a goal. 

For this, we use the definition of goals, the probability distribution over data commands $P(dc | goal)$, and incorporate it into the loss function. In order to enforce this, we introduce a Kullback-Leibler divergence term \cite{kullback1997information} into the loss function. The KL divergence measures how much a probability distribution $P$ differs from a second distribution $Q$ and has an information theoretic interpretation. Minimizing the KL divergence means optimizing the probability distribution $Pr(\hat{dc} = {dc}_{m} \vert c_{0_m}, c_{1_m}, \dots, c_{L_m}, G_m)$ to be close to the data command distribution of the goal $Pr(dc | goal)$. 

The distributions of interest in computing KL divergence are as follows
\begin{equation}
\label{equ:klloss}
\mathcal{{L_{\textrm{KL}}}}(\theta) = \mathcal{D}_{\textrm{KL}}(P \vert\vert Q)
\end{equation}
where $P = Pr(\hat{dc} = {dc}_{m} \vert c_{0_m}, c_{1_m}, \dots, c_{L_m}, G_m)$ and $Q$ $=$ $Pr(dc | goal)$.
We want our model to give recommendations that have a high $Pr(dc | goal)$ when the input from the user deviates from the chosen goal. Cross-entropy loss penalizes prediction deviating from the ground truth command. To penalize deviation from the goal, we introduce a loss function, $\mathcal{L}(\theta)$, that takes care of both the penalties:
\begin{equation}
    \label{equ:kldiv}
    \mathcal{L}(\theta) = \alpha \mathcal{L_{\textrm{CE}}}(\theta) + (1-\alpha) \mathcal{{L_\textrm{KL}}}(\theta)
\end{equation}
The first term is a scaled version of the cross-entropy loss. The second term is the KL Divergence between the predicted probability distribution and the chosen goal distribution and $\alpha$ is the balancing factor for both the losses. 


\subsection{Fine Tuning}
\label{finetuning}
The loss function described in equation \ref{equ:kldiv}, when applied to the models proposed in Section \ref{iimodels}, steers the recommender systems to produce data commands relevant to the goal under consideration. The proposed loss function involves the component $Q$, which is different for each goal. Therefore, it is not possible to train a single global model for all the goals using this loss function. A simple solution is to train $K$ different models for $K$ different goals, similar to what we proposed in the ensemble approach in Section \ref{ismodels}. 
However, models trained on large volume of diversified data from different goals overcome the problem of overfitting \cite{taigman2014deepface}. For this reason, the goal informed models, when trained on global data using the loss function $\mathcal{L}_{\textrm{CE}}(\theta)$, have an inherent advantage of learning to represent the sequences better compared to the goal specific data models. These representations are captured through the embeddings, the hidden state weights of the LSTM layer, and the fully connected layer parameters of these models. Therefore, we utilize the trained weights of these models for initializing the parameters of this experiment. We then fine tune the models to be goal specific by retraining \cite{pan2009survey} the models on the data specific to this goal with the modified loss function $\mathcal{L}(\theta)$. Consequently, these fine tuned models were able to provide accurate and goal relevant data command recommendations. Another indispensable advantage of these models is their superior performance for low resource goals, that is, goals with less training data.
\section{Experiments}
Limited research has been carried out in identifying and incorporating goal information for predicting next action in a sequence. Therefore, in this section, we compare our proposed models with simple, yet powerful, approaches used for next action prediction such as popularity based prediction models and Markov models\footnote{State-of-the-art probabilistic sequence prediction models cannot be tested since the data commands are not aligned with the actual sequence (ground truth label may occur further down the actual sequence). A comparison against neural network approach as baseline is presented in Table \ref{tab:accuracy}. }. 
\label{experiments}

\subsection{Baseline Approaches}
We provide a brief outline of the competing methods in the following.

\begin{itemize}
    \item \textbf{Top 50 Frequency Model}\quad The probability distribution, of the data commands is observed to follow Zipf's law \cite{zipf1949human}. Few data commands have a high probability of being observed in the dataset. With this observation, we modeled this baseline approach to predict only the 50 most frequent data commands. These 50 data commands constitute more than 90\% occurrence of the total number of data commands in the dataset. The corresponding probability distribution is generated by normalizing over these top 50 data commands to a unit vector.
    
    \item \textbf{Markov Models}\quad The traditional formula for computing the probability of observing a data command, $dc_i$, given the most recent command $c_L$ is:
    \begin{equation}
        \label{equ:Markov}
        Pr(\hat{dc} = dc_i \vert c_L) = \frac{count(c_L, dc_i)}{count(c_L)}
    \end{equation}
    where $count(c_L)$ is the frequency of command $c_L$ in the dataset and $count(c_L, dc_i)$ equals the frequency of the \textit{biterm} $(c_L, dc_i)$. It is to be noted that the term \emph{biterm} is used instead of \emph{bigram} in standard Markov model. Generating biterms from the dataset is explained through the following example. For a given sequence $(dc_1, dc_2, sc_1, sc_2, dc_3, sc_3, dc_4)$, the biterm set is $\{(dc_1, dc_2),$ $(dc_2,$ $dc_3),$ $(sc_1, dc_3),$ $(sc_2, dc_3),$ $(dc_3,$ $dc_4),$ $(sc_3, dc_4)\}$, where $sc_i\in \textrm{SC}$ and $dc_i\in \textrm{DC}$. In generating the biterms for the first-order Markov model, the second term in the biterm is always from the set $\textrm{DC}$, whereas the first term can be from either command sets.
    
    \item \textbf{CPT+}\quad  This is an incremental and easily adaptable approach for a lossless compression of the training data so that all relevant information is available for each prediction \cite{gueniche2015cpt+}. It relies on a tree structure and a more complex prediction algorithm to offer considerably more accurate predictions than many state-of-the-art prediction models. 
    
    \item \textbf{Vanilla Model ({\sc lstm4rec})}\quad The architecture of vanilla model (Figure \ref{fig:incore_incomm_archi}) is similar to the proposed goal informed models in Section \ref{iimodels} with no goal information being provided. It is a deep neural seq2seq model trained on entire data to predict the next data command in the sequence. This approach models user sessions with the help of an RNN with LSTM reminiscent to the one proposed in {\sc gru4rec}~\cite{hidasi2015session} by replacing the Gated Recurrent Units with LSTM units.
\end{itemize}

\subsection{Goal Identification}

The first task in identifying the goals from the dataset is to choose the optimal number of goals for the dataset at hand. We have utilized the measures described in Section \ref{coherence} to decide upon this number. We computed goal coherence scores $CS_\text{\scriptsize{\emph{UCI}}}(t)$ and $CS_\text{\scriptsize{\emph{UMass}}}(t)$ by iterating the number of goals from $t=1$ through $50$ to determine the optimal number of goals. Through a comparative analysis across these scores for various values of $t$, the value $t=6$ had the highest average coherence score. For goal identification using BTM, standard Gibbs sampling \cite{Gibbs} is used to compute the values of the multinomial distributions. The choice of the hyperparameters $\alpha$ and $\beta$ is based upon the desired output distribution of BTM. The value of hyperparameter $\alpha$ is set to 8.333 and $\beta$ is set to 0.005. 

\subsection{Model Configurations}
The length of each session, $S$, is chosen to be 30 (average command sequence length). 
Each command $c_i$ is represented as a $200$ dimensional vector in the embedding space. 
The user is asked to choose one of the goals from the list of the identified goals from the dataset at the start of each session. 
The goal is represented as the same $200$ dimensional vector for \emph{GAIn} unlike the \emph{GComm} and \emph{GCoRe} models, where it is represented as a one-hot vector. 
The number of hidden LSTM units is set to $500$ to generate the representation of the input sequence, followed by a fully connected layer. 
We further utilize dropout probability of $0.5$ in the dense layer. During training, we minimize the cross-entropy loss $\mathcal{L}_{\textrm{CE}}(\theta)$ between the logits, the predicted probability distribution over the entire data commands, and the ground truth distribution. The optimizer used is a stochastic gradient descent (SGD) with 0.01 learning rate, 0.9 momentum and weight decay of 0.001 to train our models. The weights of these models are initialized randomly with no prior distribution. 
For the fine tuning experiments, the weights are initialized to the trained values of \emph{GCoRe, GComm, GAIn} models. During fine tuning, we set the embeddings of the commands to be fixed after modifying the loss function from $\mathcal{L}_{CE}(\theta)$ to $\mathcal{L}(\theta)$. 
\section{Evaluation and Discussion}
We start this section by evaluating the identified goals using a human evaluator. We then proceed to compare the performance of the various models proposed with the competing baselines. We also propose a novel evaluation method to measure the degree of goal orientation and report the performance of our models in adversarial settings. 
\label{evaluation}
\subsection{Goal Coherence}
We recall that each goal is a probability distribution over the same set of data commands. To evaluate the quality and coherence of identified goals by BTM, which was trained in an unsupervised fashion, we rely on the assessment of one expert who had several years of experience with the software under consideration. This expert was shown top 50 commands for each of the 6 goals and was able to relate to each collection of commands and provide a phrase describing each goal. These descriptions read as follows: product analysis for campaign or marketing, search trends for campaigning research, product conversions on different devices, navigational research and funnel analysis, segment analysis for campaigning research, and search trends for web-product analysis.
\begin{figure*}[t]
  	\centering
    \includegraphics[scale=0.28]{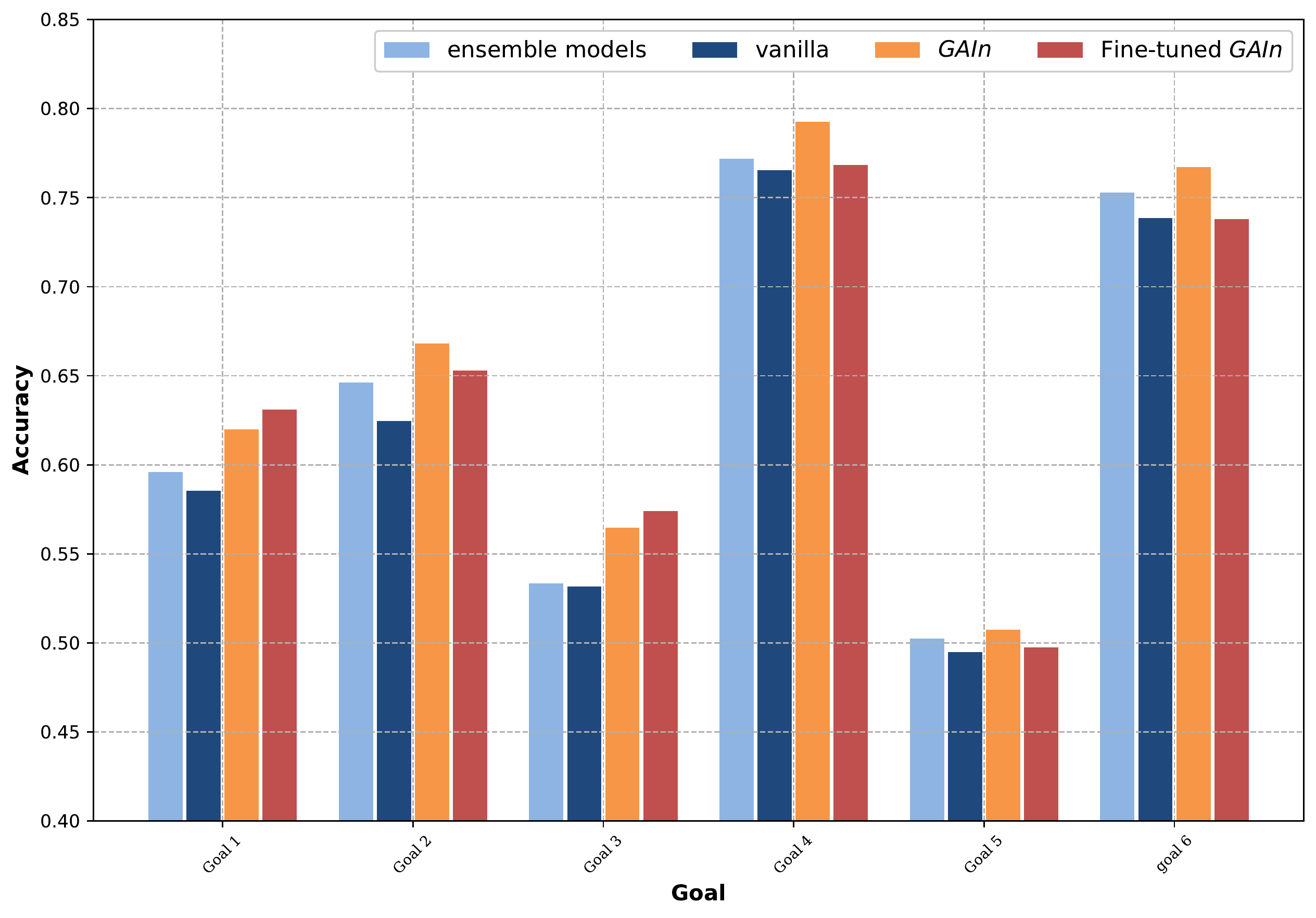}
    \hspace{4mm}
    \includegraphics[scale=0.28]{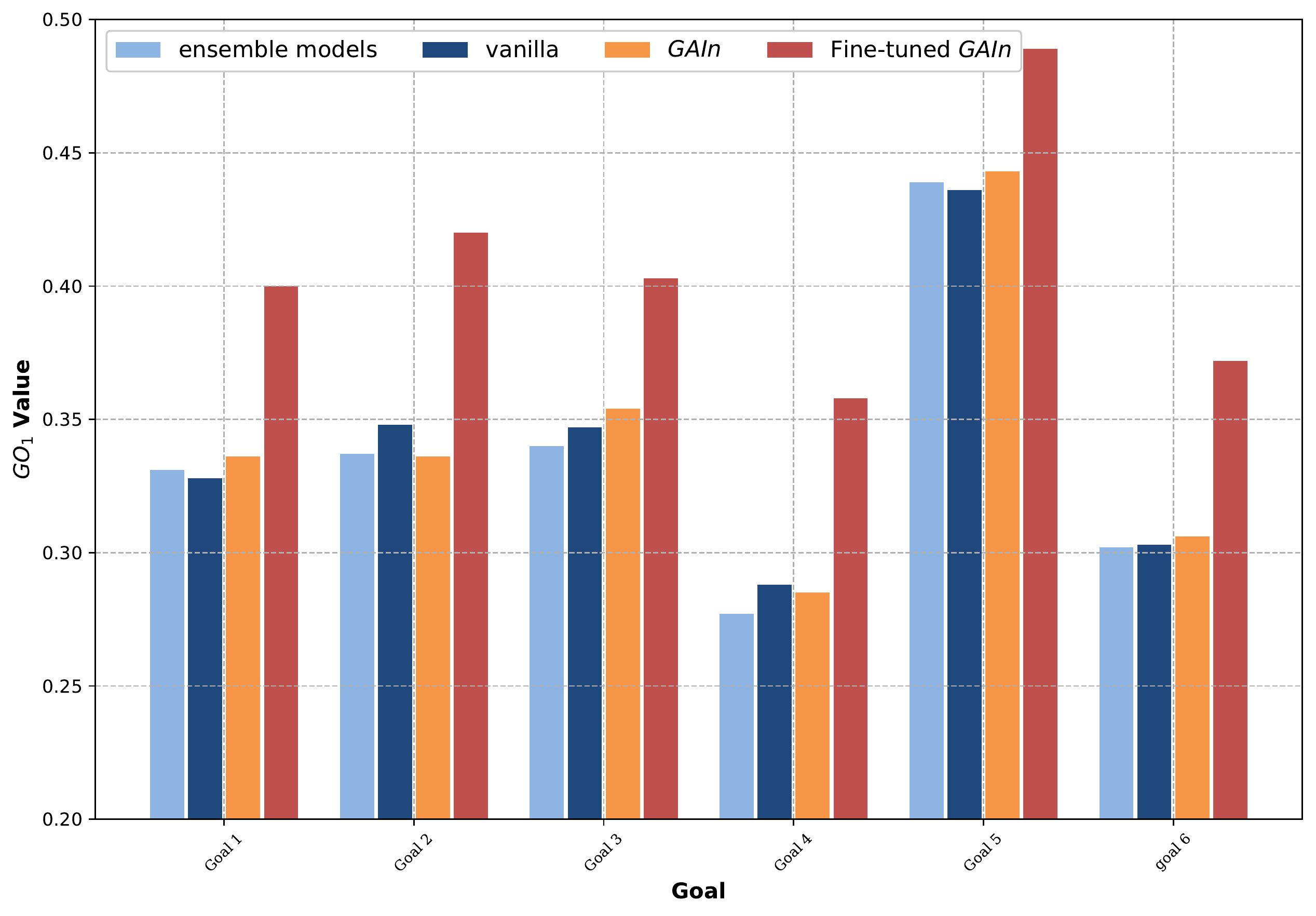}
   \caption{Accuracy values (left) and Goal orientation $GO_1$ scores (right) of proposed neural architecture models for each goal. The reported values are for the LSTM version of the models. \emph{GAIn} model outperforms for 4 out of the 6 identified goals in terms of Accuracy. The fine-tuned \emph{GAIn} model has a significant improvement in $GO_1$ score compared to other models.}
   \label{fig:plots}
\end{figure*}

\subsection{Accuracy of Models}
To quantify the performance of our models, we computed the standard evaluation metric, the test accuracy. In our setting, the test accuracy is defined as the number of examples where the predicted data command matches with the target data command out of the total number of examples. This is done at the granularity of each command in the ground truth. For uniformity, the test set across all the models was exactly the same.

From the accuracy values, it is empirically observed that the models incorporating the goal information, in general, perform better by at least 13\% margin compared to the competing baseline {\sc lstm4rec} for session based recommendations without goal information. Goal information influences the process of data command recommendation to make it more germane which is clearly reflected from the results reported in Table \ref{tab:accuracy}. 

The three Goal Informed models introduced in Section \ref{iimodels}, \emph{GCoRe}, \emph{GComm} and \emph{GAIn}, achieve similar accuracy values. The \emph{GAIn} model performed best among the three models as showcased in Table \ref{tab:accuracy}. The goal-wise accuracy results are plotted in Figure \ref{fig:plots}. The fine-tuned models also have obtained similar accuracy values with Fine-tuned {\emph{GAIn}} having best performance as observed from Table \ref{tab:finetune}. 
\begin{table}
\centering
\scalebox{0.7}{
\begin{tabular}{ c c c c }
\toprule
\multirow{2}{*}{\textsc{Model}} & \multirow{2}{*}{\textsc{Accuracy}} & \multirow{2}{*}{\textsc{$GO_1$ Score}} & \textsc{Mode of Goal Input}\\
& & & \textsc{Specification} \\
\midrule
Top 50 Frequency & 0.1633 & 0.1997 & \multirow{5}{*}{Not Specified}\\
First-order MM & 0.2621 & 0.2251 \\
Second-order MM & 0.3210 & 0.2322 \\
CPT+ & 0.3444 & 0.2542 & \\
vanilla ({\sc lstm4rec}) & 0.5875 & 0.3176 \\
\midrule
ensemble First-order MM & 0.3043 & 0.2571 & \\
ensemble Second-order MM & 0.3429 & 0.2679 & Implicitly through\\
ensemble CPT+ & 0.4154 & 0.2723 & data partition\\
ensemble Vanilla ({\sc lstm4rec}) & 0.6894 & 0.3211 &  \\
\midrule
\emph{GCoRe} & 0.6839 & \cellcolor{mygray}0.3984 & \multirow{2}{*}{Encoded}\\
\emph{GComm} & 0.6970 & 0.3959 \\
\emph{GAIn} & \cellcolor{mygray}0.7189 & 0.3933 & Representation\\
\bottomrule
\end{tabular}
}
\vspace{4mm}
\caption{Accuracy results and Goal Orientation $GO_1$ scores of various models, with best results in gray background. The ensemble models corresponds to training one model per goal.}
\label{tab:accuracy}
\vspace{-3mm}
\end{table}
\subsection{Goal Orientation Measure (\emph{GO-Measure})}
Due to the very nature of the problem we attempt to solve, conventional evaluation metrics such as test accuracy will not evaluate the effectiveness of our model, as to what extent the problem is solved. Accuracy does not convey how well the recommendations were aligned with the selected goal. 
To bring in the notion of goal awareness in the evaluation as well, we introduce the goal awareness score: the probability of the recommendation given the goal $P(dc | goal)$. Since the objective is to evaluate the model both in terms of goal awareness score and accuracy, 
we utilize a measure similar in spirit to the F-Measure \cite{sasaki2007truth}, which we term as the \emph{GO-measure} and is defined as:
\begin{equation}
    \label{equ:fbeta}
    GO_1(goal) = 2 \cdot \frac{accuracy \cdot \overline{P}(dc|\textrm{goal})}{(accuracy) + \overline{P}(dc|\textrm{goal})}
\end{equation}
where $\overline{P}(dc\vert goal)$ is the average value of $P(dc\vert goal)$ for the predicted data command $dc$ across the test examples for the specified goal. 

The $GO_1$ scores of the proposed models and fine-tuned models are reported in Tables \ref{tab:accuracy} and \ref{tab:finetune} respectively, with goal-wise results plotted in Figure \ref{fig:plots}. The $GAIn$ model performs better compared to other models when trained on loss functions $\mathcal{L}_{CE}(\theta)$ as well as $\mathcal{L}(\theta)$. The improved values of $GO_1$ reported in Table \ref{tab:finetune}, after fine-tuning the models, exhibit the recommended data commands being goal aligned with accuracy being intact. Results point that we can obtain significant improvements in terms of $GO_1$ through fine-tuning, i.e., at least 12.7\% indicating that this is very a promising direction to pursue further investigations.
\begin{table*}[h!]

\centering
\begin{tabular}{c c c}
\toprule
\textsc{Model} & \textsc{Accuracy} & \textsc{$GO_1$ Score}\\
\midrule
Vanilla & 0.5563 & 0.3852\\
Fine-tuned \emph{GCoRe} & 0.6647 & 0.4322\\
Fine-tuned \emph{GComm} & 0.6820 & 0.4268\\
Fine-tuned \emph{GAIn} & \cellcolor{mygray}0.7010 & \cellcolor{mygray}0.4811\\
\bottomrule
\end{tabular}
\vspace{4mm}
\caption{Accuracy results and Goal Orientation $GO_1$ scores of fine-tuned models with and without goal information. The fine-tuning approach is always beneficial.}
\label{tab:finetune}
\vspace{-9mm}
\end{table*}
\subsection{Adversarial testing}
Models trained on data specific distribution often display aberrant behavior when tested on an example from a different data distribution. For instance, that the user might deviate from the specified goal while progressing the session. In such cases, it is expected from the fine-tuned models to recommend data commands related to the specified goal and bring back the user on track. In this section, we test our models in such cases and display the robustness of the fine-tuned models. 

\begin{figure}
\centering
  \includegraphics[width=8.0cm,height=5.0cm]{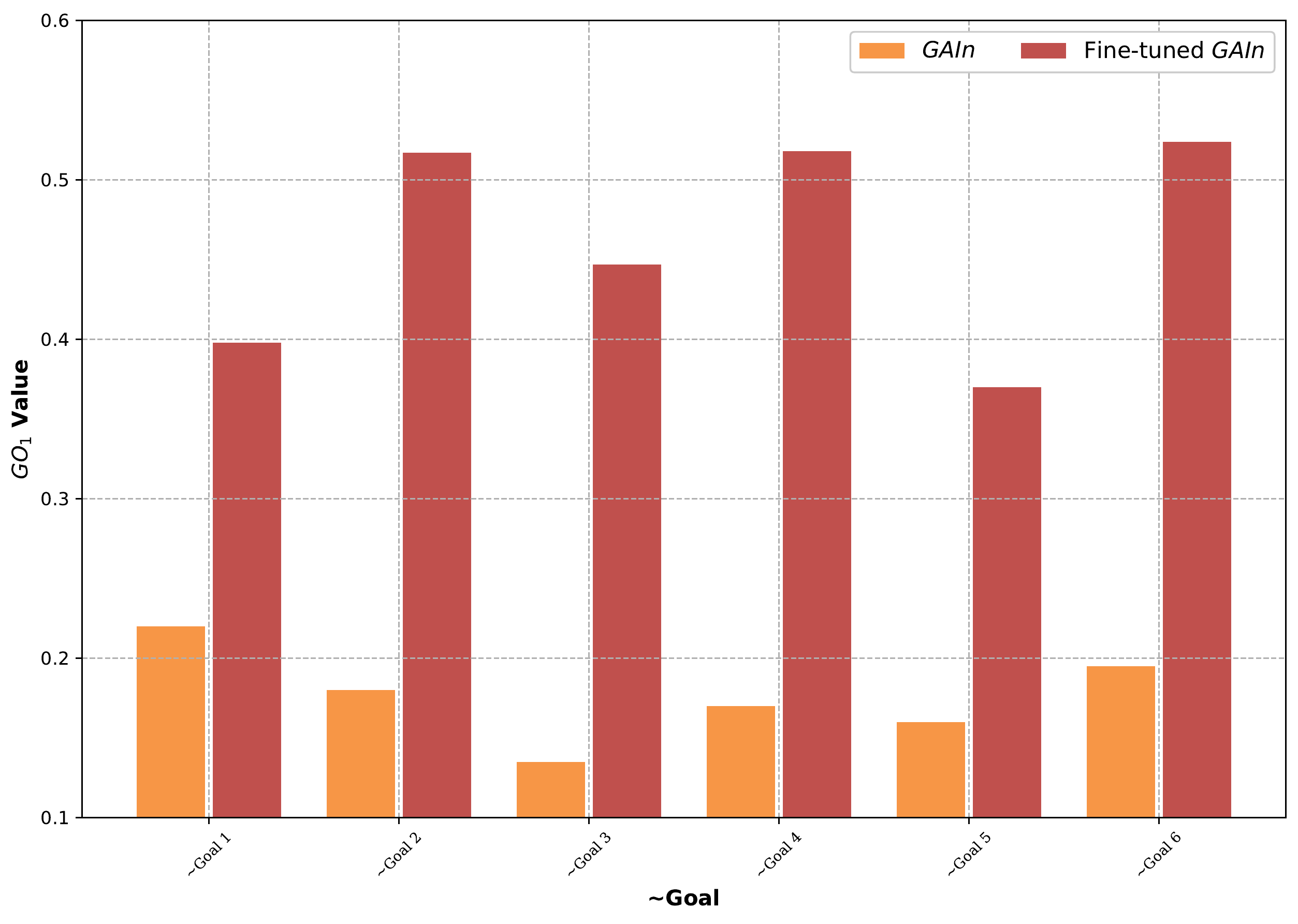}
  \caption{Goal-wise Goal Orientation $GO_1$ scores of \emph{GAIn} and fine-tuned \emph{GAIn} models on adversarial examples. The fine-tuned \emph{GAIn} model significantly outperforms \emph{GAIn} for all the goals. Note: The $\sim$Goal G on x-axis implies the model fine-tuned for goal G and tested on data from rest of the goals.}
  \vspace{-2mm}
  \label{fig:advresults}
\end{figure}
For this experiment, the model is provided inputs from data distributions different from what it was trained on. For each goal, the specific fine-tuned model, datapoints corresponding to the other goal are provided as input. The results are shown in Table \ref{tab:adversarial}. Though the accuracy of the models has decreased, the goal awareness score is intact. This provides evidence that the model is attempting to recommend data commands that are more relevant to the goal at hand. The decrease in accuracy can be interpreted as follows. The fine-tuned models were trained to predict the accurate data command for a given sequence, and then were altered to tune the recommendations towards a specific goal. When the user deviates from the specified goal, the sequence of commands might not map to that goal. Such an example, when visualized in the latent representation space of the sequences, lies outside the goal. Due to this reason, recommendations provided might not align with the original input sequence. 
However, the suggested data command steers the user back towards the intended goal.
\begin{table}
\centering
\begin{tabular}{c c c}
\toprule
\textsc{Model} & \textsc{Accuracy} & \textsc{$GO_1$ Score}\\
\toprule
ensemble vanilla & 0.1525 & 0.2566\\
\emph{GAIn} & \cellcolor{mygray}0.4919 & 0.1966\\
Fine-tuned \emph{GAIn} & 0.2795 & \cellcolor{mygray}0.4823\\
\bottomrule
\end{tabular}
\vspace{4mm}
\caption{Accuracy results and Goal Orientation $GO_1$ scores of the proposed models when tested on adversarial examples. The proposed $GO_1$ measure decreases sharply for the \emph{GAIn} model when trained on $\mathcal{L}_{CE}(\theta)$ loss function compared to the fine-tuned \emph{GAIn} model, which is trained on $\mathcal{L}(\theta)$ loss.}
\label{tab:adversarial}
\vspace{-3mm}
\end{table}
\section{Conclusion and Future work}
In this paper, we have investigated the effectiveness of incorporating goal information while recommending data commands to the user. The results of our goal-aware data command recommendation models, when compared to traditional goal-agnostic recommendation models, are quite promising and call for further explorations along this line of work. To the best of our knowledge, this is the first work which applies the fine-tuning paradigm to tailor the data commands recommendations towards the specified goal. We validate our models on a novel evaluation measure that balances both accuracy and degree of goal orientation in the provided data command recommendations. 

In future work, we would like to explore the recent advances in attention mechanisms and transfer learning to further boost the performance of our models. Also, we will experiment with more sophisticated models that can predict the user's goal in real time based on the progress of the session, which currently is an input signal given by user at the start of the session. We believe that this line of work can be extended to handle the problem of a novice user mis-specifying goal thereby providing better user experience. 

\bibliographystyle{ACM-Reference-Format}
\bibliography{sample-sigconf}

\end{document}